\documentclass[a4paper,12pt]{article}
\usepackage{amssymb,amsmath}
\usepackage{fancyhdr,appendix,verbatim}
\usepackage{parskip}
\long\def\del#1\enddel{}
\newcommand{\sect}[1]{\setcounter{equation}{0}\section{#1}}

\numberwithin{equation}{section}

\def\arccosh{\mathrm{arccosh}}

\newcommand{\tr}{\mbox{Tr}}

\def\axs{{\rm AdS}_5\times S^5}
\newcommand{\eq}[1]{\begin{equation} #1 \end{equation}}
\newcommand{\al}[1]{\begin{align} #1 \end{align}}

\begin{document}
\begin{titlepage}
\markright{\bf TUW--13--17}
\title{On the algebraic curves for circular and folded strings in AdS$_5\times S^5$}
\author{D.~Arnaudov${}^{\star}$\!, R.~C.~Rashkov${}^{\dagger,\star}$\thanks{rash@hep.itp.tuwien.ac.at}\, and T.~Vetsov${}^{\star}$
\ \\ \ \\
${}^{\star}$ Department of Physics, Sofia University,\\
5 J. Bourchier Blvd, 1164 Sofia, Bulgaria
\ \\ \ \\
${}^{\dagger}$ Institute for Theoretical Physics,\\
Vienna University of Technology,\\
Wiedner Hauptstr. 8-10, 1040 Vienna, Austria
}
\date{}
\end{titlepage}

\maketitle
\thispagestyle{fancy}

\begin{abstract}
There have been recent advances in the construction of algebraic curves for certain classes of string solutions in the context of the AdS/CFT correspondence. In this paper we obtain the Lax operators and associated spectral curves for circular and folded string solutions in AdS$_5\times S^5$. In addition, we provide an original approach for the reconstruction of string solutions in $S^3$ from their corresponding curves.
\end{abstract}

\sect{Introduction}

The AdS/CFT correspondence \cite{Maldacena:1997re,Gubser:1998bc,Witten:1998qj} is a magnificent duality which in its original setup relates $\mathcal{N}=4$ supersymmetric Yang--Mills theory and string theory in AdS$_5\times S^5$ spacetime. The string sigma model in this background is shown to be completely integrable, at least classically \cite{Minahan:2002ve,Bena:2003wd}. Integrability gives us diverse ways to find exact solutions of superstring theory at the classical level \cite{Frolov:2003qc,Kazakov:2004qf,Kazakov:2004nh}.\footnote{For an extensive review see \cite{Tseytlin:2010jv}.} The methods developed for solving integrable systems~\cite{Babelon:2003} are also powerful and efficient tools for the classification of solutions \cite{Arutyunov:2009ga,Beisert:2010jr}. The Lax formalism can be used to encode important information about the theory in the form of Riemann surfaces known as spectral curves \cite{Beisert:2004ag}--\cite{SchaferNameki:2010jy}. Their moduli contain the conserved quantities for the corresponding solution. Thus, instead of analyzing string solutions, we can use the analytical properties of these Riemann surfaces to examine different aspects of the theory.

The construction of algebraic curves for closed-string solutions has been based on analysis of the monodromy around a non-contractible loop going around the closed-string cylinder \cite{Beisert:2005bm,SchaferNameki:2010jy}. The spectral curve has been defined through a Lax operator, which is the derivative with respect to the spectral parameter of the logarithm of the monodromy operator \cite{Dorey:2006zj,Sakai:2006bp,Zarembo:2010yz}. The quasimomenta for the full $\axs$ superstring in light-cone gauge have been obtained in \cite{Alday:2005gi}. The authors of \cite{Gromov:2007aq} have explicitly derived the quasimomenta for circular strings in $\axs$. In \cite{Gromov:2011de} a similar analysis has been provided for the case of the folded string with two charges. The full quantum spectral curve has been constructed within the same approach in \cite{Gromov:2013pga}. Related to the finite-gap method in \cite{Kazakov:2004qf} is the construction of Wilson loops with the help of theta functions \cite{Ishizeki:2011bf}. Recently the algebraic curves for long folded and circular strings in AdS$_5\times S^5$, the null cusp Wilson loop and $q\bar q$ potential have been studied \cite{Janik:2012ws,Ryang:2012uf}, providing important information about the theory.

However, there is a local construction presented in \cite{Dekel:2013dy}, which does not require the complicated analysis of the asymptotics imposed on the solutions and the monodromy matrix at all. It simplifies considerably the calculation of the Lax operator and can be implemented for certain classes of solutions to sigma-models on group manifolds where one can represent the dependence of the flat connection on one of the worldsheet variables as a similarity transformation with a matrix $S(\tau)$ satisfying $S^{-1}dS={\rm const}$, i.e. $A(\tau,\sigma)=S^{-1}(\tau)A_0(\sigma)S(\tau)$, or respectively $A(\tau,\sigma)=R^{-1}(\sigma)\tilde A_0(\tau)R(\sigma)$. This is called a factorized flat connection. If the model allows both factorizations, we have a complete factorization. As it turns out, almost all interesting for AdS/CFT correspondence string solutions in AdS$_5\times S^5$ admit a factorized flat connection. Some examples have already been considered in \cite{Dekel:2013dy}, in particular the curve for the folded spinning string in AdS$_3$ \cite{Kazakov:2004nh,Zarembo:2010yz}.

In this paper, using the approach presented in \cite{Dekel:2013dy}, we will construct the spectral curves for circular and folded string solutions in AdS$_5\times S^5$. In Section \ref{sec2} we will discuss briefly the method. In Section \ref{sec3} we will consider the case of circular strings in AdS$_3\times S^3$. Next we will examine the curve for the folded string in $S^3$. In Section \ref{sec5} we will analyse the algebraic curves for circular spinning strings in AdS$_5$ and $\mathbb{R}\times S^5$. After that we will study the corresponding curves for various rigid strings in AdS$_5$. We will conclude with a brief discussion on the results in Section \ref{sec7}. A procedure for the reconstruction of string solutions in AdS$_3$ and $S^3$ from their spectral curves will be presented in an Appendix.

\sect{General setup}\label{sec2}

The main goal of this paper is the analysis of classical string solutions through their corresponding spectral curves. We will investigate the dynamics of strings in AdS$_5$, $S^5$, and their submanifolds AdS$_3$ and $S^3$. These spaces are either group or coset manifolds. Therefore they can be parameterized by a group element $g\in G$, belonging to a certain Lie group $G$. Consequently, the starting point of our analysis is the worldsheet sigma-model action
\eq{
S=\int d^2\sigma\,\tr(j\wedge\ast j)\,,
\label{sigma}
}
where the current $j\equiv g^{-1}dg$ is the Maurer--Cartan (MC) one-form taking values in the algebra $\mathfrak{g}$ of $G$ \cite{Bena:2003wd}. The worldsheet current $j$ enters the connection $A$ in the following manner\footnote{We use a worldsheet metric with Minkowskian signature throughout the paper.}
\eq{
A=\frac{1}{1-z^2}\left(j-z\ast\!j\right)=\frac{1}{1-z^2}\,(j_\tau+zj_\sigma)\,d\tau+\frac{1}{1-z^2}\,(j_\sigma+zj_\tau)\,d\sigma\,,
\label{Afromj}
}
where we have assumed that the Hodge dual acts as $\ast\,d\tau=-d\sigma$ and $\ast\,d\sigma=-d\tau$. We will work with string solutions admitting a factorized connection, which means that $A$ can be expressed either as \cite{Dekel:2013dy}
\begin{equation}
A(\tau,\sigma)=S^{-1}(\tau)A_0(\sigma)S(\tau)\,,\quad S(\tau)\in G\,,\quad A_0(\sigma)\in\mathfrak{g}\,,
\end{equation}
when we have a connection factorized with respect to $\tau$, or
\begin{equation}
A(\tau,\sigma)=R^{-1}(\sigma)\tilde A_0(\tau)R(\sigma)\,,\quad R(\sigma)\in G\,,\quad\tilde A_0(\tau)\in\mathfrak{g}\,,
\end{equation}
when $A$ is factorized with respect to $\sigma$. It is useful to define for later use the following left and right currents $j_S^L\equiv S^{-1}dS={\rm const}$, $j_S^R\equiv dSS^{-1}={\rm const}$, and correspondingly $j_R^L\equiv R^{-1}dR={\rm const}$, $j_R^R\equiv dRR^{-1}={\rm const}$.

The equations of motion following from \eqref{sigma} constrain the connection $A$ to be flat, i.e. it obeys the condition
\eq{
dA+A\wedge A=0\,.
}
Taking into account that $A$ is factorized, this leads to the following equation in the case of factorization with respect to $\tau$
\begin{equation}
\partial_\sigma A_{0,\tau}(\sigma)=[A_{0,\tau}(\sigma)-j_S^R,A_{0,\sigma}(\sigma)]\,.
\label{tauflat}
\end{equation}
When we have factorization with respect to $\sigma$, we obtain
\begin{equation}
\partial_\tau\tilde A_{0,\sigma}(\tau)=[\tilde A_{0,\sigma}(\tau)-j_R^R,\tilde A_{0,\tau}(\tau)]\,.
\label{sigmaflat}
\end{equation}
By virtue of \eqref{tauflat} and \eqref{sigmaflat} we can define a Lax operator as a rational function of the spectral parameter $z\in\mathbb{C}$
\begin{equation}
L(\tau,\sigma)\equiv A_\tau(\tau,\sigma)-j_S^L
\end{equation}
for the case of $\tau$-factorization, or
\begin{equation}
L(\tau,\sigma)\equiv A_\sigma(\tau,\sigma)-j_R^L\,,
\end{equation}
when we have $\sigma$-factorization. The above Lax operators can be straightforwardly shown to satisfy the equations
\begin{equation}
\partial_\tau L=[L,A_\tau]\,,\qquad\partial_\sigma L=[L,A_\sigma]\,.
\end{equation}
The associated spectral curve $\Gamma$ is defined through the characteristic equation for the eigenvalues of the Lax matrix \cite{Babelon:2003,Janik:2012ws,Dekel:2013dy}
\begin{equation}
\det(L-y\mathbf{1}_{n\times n})=0\,,\qquad y\in\mathbb{C}\,,
\label{curve}
\end{equation}
where $n$ is the dimension of the fundamental representation of $\mathfrak{g}$. The solution to \eqref{curve} is an algebraic curve in $\mathbb{C}^2$. One can prove that
\begin{equation}
\det(L-y\mathbf{1}_{n\times n})\propto\frac{1}{(1-z^2)^n}\sum_{i=0}^{2n}c_i(y)z^i,
\end{equation}
where $c_i(y)$ are polynomials in $y$ of order $n$. For the simplest case we are interested in, namely $n=2$, \eqref{curve} assumes the form
\eq{
y^2=-\det(L)=\frac12\tr(L^2)=\frac12\frac{1}{(1-z^2)^2}\sum_{i=0}^{4}c_iz^i,
\label{curvepoly}
}
where we have used that for the backgrounds of interest the algebras $\mathfrak{g}$ consist of traceless matrices, which also leads to the curve being independent of $\tau$ and $\sigma$.

In order to construct the Lax operator and the spectral curve it is much easier to work with the corresponding MC equations for the currents, which can be derived in the following way. First we substitute the connection $A$ from \eqref{Afromj} in equations \eqref{tauflat} or \eqref{sigmaflat}, respectively. Then we expand the relevant equation in powers of the spectral parameter $z$. Thus for $\tau$-factorization we find\footnote{Note the sign difference between the second equation in \eqref{eqcurrents} and (2.11) of \cite{Dekel:2013dy} due to our use of Minkowskian signature for the worldsheet.}
\begin{equation}
\partial_\sigma j_\tau=[j_\tau-j_S,j_\sigma]\,,\qquad\partial_\sigma j_\sigma=-[j_S,j_\tau]\,,
\label{eqcurrents}
\end{equation}
where we have denoted $j_S^L$ by $j_S$. For the case of factorization with respect to $\sigma$ we have
\begin{equation}
\partial_\tau j_\tau=-[j_R,j_\sigma]\,,\qquad\partial_\tau j_\sigma=[j_\sigma-j_R,j_\tau]\,,\qquad j_R\equiv j_R^L\,.
\end{equation}
By solving the above equations for $j_S$ or $j_R$ and using that the flat connection $A$ follows in a straightforward way from the string solution, we can eventually obtain the Lax operator and curve. In the following sections we will apply this method to construct the spectral curves for circular and folded string solutions in AdS$_5\times S^5$.

\sect{Circular spinning string in ${\rm AdS}_3\times S^3$}\label{sec3}

The first case we will consider is the most general form of a circular rotating string in AdS$_3\times S^3$ \cite{Frolov:2003qc,Arutyunov:2003uj,Arutyunov:2003rg,Arutyunov:2003za}. The metric in global coordinates is
\eq{
ds_{{\rm AdS}_3\times S^3}^2=
-\cosh^2\rho\,dt^2+d\rho^2+\sinh^2\rho\,d\phi^2+d\theta^2+\cos^2\theta\,d\varphi_1^2+\sin^2\theta\,d\varphi_2^2\,.
}
The corresponding solution in embedding coordinates looks like
\al{\label{circsola}
Y_5+iY_0&=a_0e^{\kappa\tau},\qquad Y_1+iY_2=a_1e^{i(w\tau+k\sigma)}\,,\qquad a_0^2-a_1^2=1\,,\\
X_1+iX_2&=b_1e^{i(\omega_1\tau+m_1\sigma)}\,,\qquad X_3+iX_4=b_2e^{i(\omega_2\tau+m_2\sigma)}\,,\qquad b_1^2+b_2^2=1\,,
\label{circsols}
}
and in global coordinates
\al{
t&=\kappa\tau\,,\qquad\rho=\arccosh\,a_0=\rho_0\,,\qquad\phi=w\tau+k\sigma\,,\\
\theta&=\arccos\,b_1=\theta_0\,,\qquad\varphi_1=\omega_1\tau+m_1\sigma\,,\qquad\varphi_2=\omega_2\tau+m_2\sigma\,.
}
The following relations between parameters of the solution result from the equations of motion
\eq{
w^2=\kappa^2+k^2\,,\qquad\omega_1^2-m_1^2=\omega_2^2-m_2^2\,.
}
The Virasoro constraints give
\eq{
\kappa^2\cosh^2\!\rho_0=(w^2+k^2)\sinh^2\!\rho_0+(\omega_1^2+m_1^2)\cos^2\theta_0+(\omega_2^2+m_2^2)\sin^2\theta_0\,,
}
\eq{
wk\sinh^2\!\rho_0+\omega_1m_1\cos^2\theta_0+\omega_2m_2\sin^2\theta_0=0\,.
}
The string solution can also be characterized in terms of the following conserved charges
\eq{
E=\sqrt{\lambda}\cosh^2\!\rho_0\,\kappa\,,\ \ S=\sqrt{\lambda}\sinh^2\!\rho_0\,w\,,\ \ J_1=\sqrt{\lambda}\cos^2\theta_0\,\omega_1\,,\ \ J_2=\sqrt{\lambda}\sin^2\theta_0\,\omega_2\,.
}
The present solution lives in the direct product of group manifolds AdS$_3$ and $S^3$. The group element of AdS$_3\times S^3$ takes a block-diagonal form including the individual group elements of the product space AdS$_3\times S^3$. In order to find the corresponding algebraic curve we can work with the direct product. However, it is obvious that the curve will actually decompose to two curves: one for the AdS$_3$ part of the solution and the other for the dynamics in $S^3$. Consequently, we intend to analyse separately the two resulting curves.

\subsection{Curve for circular string in ${\rm AdS}_3$}
The spectral curve for a similar solution was found in \cite{Ryang:2012uf}. However, there the author used the approach of \cite{Janik:2012ws}, which is different from ours. The parametrization of AdS$_3$ in embedding and global coordinates is given by
\begin{equation}
g_{{\rm AdS}_3}=\left(\begin{array}{cc}
Y_5+iY_0 & Y_1+iY_2\\
Y_1-iY_2 & Y_5-iY_0\end{array}\right)=
\left(\begin{array}{cc}
e^{it}\cosh\rho & e^{i\phi}\sinh\rho\\
e^{-i\phi}\sinh\rho & e^{-it}\cosh\rho
\end{array}\right)\in{\rm SL}(2,\mathbb{R})\,.
\end{equation}
The solution in \eqref{circsola} admits complete factorization of the connection, i.e. it can be factorized with respect to both $\tau$ and $\sigma$. First, with respect to $\tau$ we obtain, using \eqref{eqcurrents}, that
\eq{
S(\tau)=e^{i(\kappa-w)\sigma_3\tau/2},\quad j_S=i(\kappa-w)\sigma_3/2\,,
}
where $\sigma_3$ is one of the Pauli matrices. The resulting curve is
\eq{
y^2=-\det(A_\tau-j_S)=-\frac{[(\kappa-w)z^2+\kappa+w]^2+4z(kz+\kappa+w)[(\kappa-w)z-k]\sinh^2\!\rho_0}{4(1-z^2)^2}\,.
\label{curvecirctau}
}
By rescaling $y$ we arrive at the canonical hyperelliptic form of the algebraic curve
\al{\label{curvecirctau'}
y^2&=\sum_{i=0}^4c_iz^i,\qquad c_0=-\frac{(\kappa+w)^2}{2}\,,\qquad c_1=2k(\kappa+w)\sinh^2\!\rho_0\,,\\
c_2&=k^2(1+4\sinh^2\!\rho_0)\,,\qquad c_3=-2k(\kappa-w)\sinh^2\!\rho_0\,,\qquad c_4=-\frac{(\kappa-w)^2}{2}\,.
\nonumber
}
The energy $E$ and angular momentum $S$ of the string can be expressed in terms of the above coefficients
\eq{
E=\sqrt{\frac{\lambda}{2}}\,\frac{(c_1-c_3)(c_2+6\sqrt{c_0c_4})}{4(c_0c_4)^{1/4}(c_2-\sqrt{c_0c_4})}\,,\quad S=\sqrt{\frac{\lambda}{2}}\,\frac{c_1+c_3}{4(c_0c_4)^{1/4}}\,.
}
Another representation of the curve is
\eq{
y^2=-[k-(\kappa-w)z]^2\!\left(\frac{ke^{2\rho_0}}{\kappa-w}+z\right)\!\left(\frac{ke^{-2\rho_0}}{\kappa-w}+z\right),
\label{ads3circ}
}
with a ramification point
\eq{
z_1=\frac{k}{\kappa-w}\,,
}
and two branch points
\eq{
z_2=-\frac{ke^{2\rho_0}}{\kappa-w}\,,\qquad z_3=-\frac{ke^{-2\rho_0}}{\kappa-w}\,.
}
The presence of the square factor is due to the complete factorization of the connection as detailed in \cite{Dekel:2013dy}. It leads also to the spectral curve being a genus-0 Riemann surface. Let us also note that $z_2$ and $z_3$ coalesce when the string is located at the center of AdS$_3$ ($\rho_0=0$), i.e. there are no branch points any more.

To constrain the parameters defining the curve we have used only the equations of motion so far. For consistency reasons let us also impose the Virasoro constraints. If we further restrict the dynamics only to the AdS$_3$ part of the background, the parameters of the solution obey the following relations
\eq{
\kappa=w\,,\qquad k=0\,,\qquad\rho_0\rightarrow+\,\infty\,,
}
which lead to the algebraic curve
\eq{
y^2=-4\kappa^2(1-z^2)=-\frac{4(E-S)^2}{\lambda}(1-z^2)\,.
}
The curve bears $(1-z^2)$ factor originating from imposition of the Virasoro constraints as discussed in \cite{Dekel:2013dy}. The spectral curve coincides with the curve for long folded strings in AdS$_3$ \cite{Dekel:2013dy}.

Let us also investigate the algebraic curve resulting from $\sigma$-factorization. In this case the similarity transformation matrix $R(\sigma)$ and the corresponding constant current can be easily obtained to be
\eq{
R(\sigma)=e^{-ik\sigma_3\sigma/2},\quad j_R=-ik\sigma_3/2\,.
}
The curve then reads off
\eq{
y^2=-\det(A_\sigma-j_R)=-\frac{(kz^2-2\kappa z-k)^2-4z(kz-\kappa+w)[(\kappa+w)z+k]\sinh^2\!\rho_0}{4(1-z^2)^2}\,,
}
or in hyperelliptic form
\al{
y^2&=\sum_{i=0}^4c_iz^i,\qquad c_0=-\frac{k^2}{2}\,,\qquad c_1=-2k\kappa-2k(\kappa-w)\sinh^2\!\rho_0\,,\\
c_2&=-2\kappa^2+k^2(1+2\sinh^2\!\rho_0)\,,\qquad c_3=2k\kappa+2k(\kappa+w)\sinh^2\!\rho_0\,,\qquad c_4=-\frac{k^2}{2}\,.
\nonumber
}
We can also represent the algebraic curve as
\eq{
y^2=-[\kappa-w-kz]^2\!\left(\frac{ke^{2\rho_0}}{\kappa-w}+z\right)\!\left(\frac{ke^{-2\rho_0}}{\kappa-w}+z\right),
}
which resembles \eqref{ads3circ} as expected.

\subsection{Curve for circular string in $S^3$}
Our next task is the calculation of the spectral curve for circular rotating strings in $S^3$ following the same line of consideration. It is useful to have the parametrization of $S^3$ in embedding and global coordinates. In this particular case it can be expressed as
\begin{equation}
g_{S^3}=\left(\begin{array}{cc}
X_1+iX_2 & X_3+iX_4\\
-X_3+iX_4 & X_1-iX_2\end{array}\right)=
\left(\begin{array}{cc}
e^{i\varphi_1}\cos\theta & e^{i\varphi_2}\sin\theta\\
-e^{-i\varphi_2}\sin\theta & e^{-i\varphi_1}\cos\theta
\end{array}\right)\in{\rm SU}(2)\,.
\end{equation}
We will work out the case of the circular string solution given by the ansatz \eqref{circsols}. The solution again admits complete factorization. The factorization with respect to $\tau$ leads to
\eq{
S(\tau)=e^{i(\omega_1-\omega_2)\sigma_3\tau/2},\quad j_S=i(\omega_1-\omega_2)\sigma_3/2\,,
}
where $\sigma_3$ is the third Pauli matrix. After rescaling $y$ we obtain for the resulting spectral curve
\eq{
y^2=4z[\omega_1+\omega_2+(m_1+m_2)z][(\omega_1-\omega_2)z+m_1-m_2]\sin^2\theta_0-[\omega_1+\omega_2+2m_1z+(\omega_1-\omega_2)z^2]^2.
\label{s3circt}
}
It can be seen that again the curve is hyperelliptic, i.e. it has the form $y^2=\sum_{i=0}^4c_iz^i$. Another representation of the algebraic curve is
\eq{
y^2=-[m_1-m_2+(\omega_1-\omega_2)z]^2\!\left(\frac{(m_1+m_2)e^{2i\theta_0}}{\omega_1-\omega_2}+z\right)\!\left(\frac{(m_1+m_2)e^{-2i\theta_0}}{\omega_1-\omega_2}+z\right),
}
with a ramification point
\eq{
z_1=-\frac{m_1-m_2}{\omega_1-\omega_2}\,,
}
and branch points
\eq{
z_2=-\frac{(m_1+m_2)e^{2i\theta_0}}{\omega_1-\omega_2}\,,\qquad z_3=-\frac{(m_1+m_2)e^{-2i\theta_0}}{\omega_1-\omega_2}\,.
}
The square factor follows from the complete factorization of the connection as detailed in \cite{Dekel:2013dy}. Its presence means that the spectral curve is a genus-0 Riemann surface. In the special case of dynamics only in $S^1\subset S^3$, which translates into $\theta_0=0$ and $\omega_2=0=m_2$, the curve reduces to
\eq{
y^2=-\omega_1^2\!\left(\frac{m_1}{\omega_1}+z\right)^4=-\frac{J_1^2}{\lambda}\!\left(\frac{\sqrt{\lambda}\,m_1}{J_1}+z\right)^4\!,
\label{circsols'}
}
i.e. all points coalesce into $-m_1/\omega_1$.

If we factorize the connection for the solution with respect to $\sigma$, the similarity transformation $R$ and the constant current will become
\eq{
R(\sigma)=e^{i(m_1-m_2)\sigma_3\sigma/2},\quad j_R=i(m_1-m_2)\sigma_3/2\,.
}
The curve can be obtained from \eqref{s3circt} by exchanging $\omega_1\longleftrightarrow m_1$ and $\omega_2\longleftrightarrow m_2$. The algebraic curve can also be represented as
\eq{
y^2=-[\omega_1-\omega_2+(m_1-m_2)z]^2\!\left(\frac{(m_1+m_2)e^{2i\theta_0}}{\omega_1-\omega_2}+z\right)\!\left(\frac{(m_1+m_2)e^{-2i\theta_0}}{\omega_1-\omega_2}+z\right).
}

We would also like to analyze a particular ansatz describing a string with equal spins
\eq{
\theta=\frac{\pi}{4}\,,\qquad\varphi_1=\omega\tau+m\sigma\,,\qquad\varphi_2=\omega\tau-m\sigma\,,
}
which automatically satisfies the equations of motion. Factorizing the $S^3$ connection with respect to $\tau$, we get $S(\tau)=\mathbf{1}\,,\ j_S=\mathbf{0}\,$. After rescaling $y$ we obtain for the resulting spectral curve
\eq{
y^2=-\omega^2-m^2z^2=-\frac{4J^2}{\lambda}-m^2z^2,
\label{s3circt'}
}
where $J=J_1=J_2$. Starting with $\sigma$-factorization, we get for the similarity transformation $R$ and the constant current
\eq{
R(\sigma)=e^{im\sigma_3\sigma},\quad j_R=im\sigma_3\,.
}
The algebraic curve is
\eq{
y^2=-(\omega^2+m^2z^2)z^2=-\!\left(\frac{4J^2}{\lambda}+m^2z^2\right)\!z^2.
\label{s3circs}
}

\sect{Spectral curve for folded spinning string in $\mathbb{R}\times S^3$}\label{sec4}

Let us also compute the algebraic curve for the folded spinning string with two angular momenta in $S^3$ \cite{Frolov:2003xy}. The corresponding ansatz in global coordinates looks like
\eq{
t=\kappa\tau\,,\qquad\theta=\theta(\sigma)\,,\qquad\varphi_1=\omega_1\tau\,,\qquad\varphi_2=\omega_2\tau\,.
}
The equation of motion for $\theta$ takes the form
\eq{
\theta''+(\omega_2^2-\omega_1^2)\,\sin\theta\,\cos\theta=0\,.
}
The solution is expressed in terms of the Jacobi amplitude function
\eq{
\theta(\sigma)={\bf am}\!\left(\sqrt{\omega_2^2-\omega_1^2-c_1}\,(\sigma+c_2)\Bigg|\frac{\omega_1^2-\omega_2^2}{\omega_1^2-\omega_2^2+c_1}\right).
\label{s3foldedsol}
}
We set $c_2=0$ by imposing that $\theta(0)=0$. Note that, strictly speaking, the string is folded only for $c_1\geq0$ and is circular otherwise. The Virasoro constraint is
\eq{
\theta'^2+\omega_1^2\,\cos^2\theta+\omega_2^2\,\sin^2\theta-\kappa^2=0\,,
}
which, given the solution \eqref{s3foldedsol}, reduces to
\eq{
c_1=\omega_2^2-\kappa^2\,.
}
The components of the MC one-form can be easily calculated, and clearly the solution is factorized only with respect to $\tau$. We obtain for the constituent $S^3$ sigma model, using \eqref{eqcurrents}, that
\eq{
S(\tau)=e^{i(\omega_1-\omega_2)\sigma_3\tau/2},\quad j_S=i(\omega_1-\omega_2)\sigma_3/2\,.
}
The resulting curve is given by
\eq{
y^2=-\det(A_\tau-j_S)=-\frac{[(1-z^2)\omega_1+(1+z^2)\omega_2]^2-4c_1z^2}{4(1-z^2)^2}\,.
}
It can be noted that in this case the algebraic curve is elliptic, i.e. of genus one. Using the Virasoro constraint, we obtain
\eq{
y^2=-\frac{(\omega_1-\omega_2)^2z^4+2(2\kappa^2-\omega_1^2-\omega_2^2)z^2+(\omega_1+\omega_2)^2}{4(1-z^2)^2}\,.
\label{foldeds3}
}
As we have mentioned above, the string is folded for $\kappa\leq\omega_2$ and circular for $\kappa>\omega_2$. At the transition ($\kappa=\omega_2$) we have
\eq{
y^2=-\frac{[(\omega_2-\omega_1)z^2+\omega_1+\omega_2]^2}{4(1-z^2)^2}\,.
}

Another noteworthy case is the circular winding string, when $\omega_1=\omega_2=\omega$, which leads to $\theta=n\sigma,\ n\in\mathbb{Z}\,,$ with $c_1=-n^2$. The Virasoro constraint becomes $\kappa^2=\omega^2+n^2$. The relevant spectral curve assumes the form
\eq{
y^2=-\frac{\omega^2+n^2z^2}{(1-z^2)^2}\,,
}
which is practically the same as \eqref{s3circt'}, because the corresponding solutions are related to each other by a global SU(2) rotation. In this particular case there are two branch points at $\pm i\omega/n$. Let us see what happens if we try to obtain the curve via $\sigma$-factorization. The constant current is $j_R=in\sigma_2$, and
\eq{
y^2=-\frac{z^2(\omega^2+n^2z^2)}{(1-z^2)^2}\,.
}
Similarly, the algebraic curve coincides with \eqref{s3circs}.

\sect{Curves for circular strings in ${\rm AdS}_5$ and $\mathbb{R}\times S^5$}\label{sec5}

In this section we will provide the spectral curves for circular spinning strings in AdS$_5$ and $\mathbb{R}\times S^5$ \cite{Frolov:2003qc,Arutyunov:2003uj,Arutyunov:2003rg,Arutyunov:2003za}. In particular limits we will recover curves that were studied in preceding sections.

\subsection{Curve for circular strings in ${\rm AdS}_5$}
The background AdS$_5$ is the coset manifold SU(2,\,2)/SO(4,\,1). The corresponding sigma model can be obtained by utilizing the following parametrization \cite{Alday:2005ww,Arutyunov:2009ga}
\begin{equation}
g_{{\rm AdS}_5}=\left(
\begin{array}{cccc}
  0  &  v_3   &  v_1   & v_2  \\
-v_3 &   0    & -v_2^* & v_1^*\\
-v_1 & v_2^*  &   0    & v_3^*\\
-v_2 & -v_1^* & -v_3^* &  0
\end{array}
\right)\in{\rm SU}(2,2)\,,
\end{equation}
\[
v_1=\sinh\rho\,\cos\psi\,e^{i\phi_1}\,,\quad v_2=\sinh\rho\,\sin\psi\,e^{i\phi_2}\,,\quad v_3=\cosh\rho\,e^{it}\,.
\]
The AdS$_5$ metric in global coordinates is given by
\eq{
ds_{{\rm AdS}_5}^2=-\cosh^2\rho\,dt^2+d\rho^2+\sinh^2\rho\,(d\psi^2+\cos^2\psi\,d\phi_1^2+\sin^2\psi\,d\phi_2^2)\,.
}
The most universal ansatz for circular strings in AdS$_5$ leads to rather complicated algebraic curve, which is in general hyperelliptic of genus three. In the transition AdS$_5\longrightarrow{\rm AdS}_3$ we recover the curve in \eqref{curvecirctau} and a similar one, obtained from \eqref{curvecirctau} by replacing $\kappa$ with $-\kappa$. In order to simplify the exposition we consider the following ansatz for a circular string in AdS$_5$ with two equal spins
\eq{
t=\kappa\tau\,,\quad\rho=\rho_0\,,\quad\psi=\frac{\pi}{4}\,,\quad\phi_1=w\tau+k\sigma\,,\quad\phi_2=w\tau-k\sigma\,.
}
The equations of motion are equivalent to
\eq{
w^2=k^2+\kappa^2\,.
}
The first Virasoro constraint gives
\eq{
\kappa^2\cosh^2\!\rho_0=(w^2+k^2)\sinh^2\!\rho_0\,,
}
while the second is automatically satisfied. The above solution admits complete factorization of the connection with
\begin{equation}
j_S={\rm diag}\left\{\frac{i}{2}(\kappa+2w),\frac{i}{2}(\kappa-2w),-\frac{i}{2}\kappa,-\frac{i}{2}\kappa\right\},
\label{ads5circjs}
\end{equation}
which leads to the following spectral curve
\al{
&y^4+\frac{\kappa^2(1-z^2)^2+2w^2(1+z^4)+4(\kappa^2-w^2)c_{2\rho_0}z^2}{2(1-z^2)^2}y^2+\frac{i\kappa w^2(1+z^2)}{1-z^2}y+\frac{(1+z^2)^2}{16(1-z^2)^4}\\
&\times\![\kappa^4(1-z^2)^2-4\kappa^2w^2(1-z^2)^2+4w^4z^2+8w^2(\kappa^2-w^2)c_{2\rho_0}z^2+4(\kappa^2-w^2)^2c_{2\rho_0}^2z^2]=0\,,
\nonumber
}
where we have denoted $c_{2\rho_0}\equiv\cosh(2\rho_0)$. Imposing the first Virasoro constraint, we get
\eq{
y^4+\frac12(\kappa^2+2w^2)y^2+\frac{i\kappa w^2(1+z^2)}{1-z^2}y+\frac{\kappa^2(\kappa^2-4w^2)(1+z^2)^2}{16(1-z^2)^2}=0\,.
\label{ads5circ}
}
The connection can also be factorized with respect to $\sigma$, in which case we obtain
\eq{
y^4+k^2y^2-\frac{i\kappa(\kappa^2+2k^2)z}{1-z^2}y-\frac{\kappa^2(3\kappa^2+4k^2)z^2}{4(1-z^2)^2}=0\,,
}
where we have used the Virasoro constraint.

\subsection{Curve for circular strings in $\mathbb{R}\times S^5$}
The five-sphere, which represents the coset manifold SU(4)/SO(5), can be parameterized by \cite{Alday:2005ww,Arutyunov:2009ga}
\begin{equation}
g_{S^5}=\left(
\begin{array}{cccc}
  0  &  u_3   &  u_1   & u_2\\
-u_3 &   0    & u_2^*  & -u_1^*\\
-u_1 & -u_2^* &   0    & u_3^*\\
-u_2 & u_1^*  & -u_3^* & 0
\end{array}
\right)\in{\rm SU}(4)\,,
\end{equation}
\[
u_1=\sin\gamma\,\cos\theta\,e^{i\varphi_1}\,,\quad u_2=\sin\gamma\,\sin\theta\,e^{i\varphi_2}\,,\quad u_3=\cos\gamma\,e^{i\varphi_3}\,.
\]
The background metric is
\eq{
ds_{\mathbb{R}\times S^5}^2=-dt^2+d\gamma^2+\cos^2\gamma\,d\varphi_3^2+\sin^2\gamma\,(d\theta^2+\cos^2\theta\,d\varphi_1^2+\sin^2\theta\,d\varphi_2^2)\,.
}
As in the AdS$_5$ case the general curve is complicated, therefore we will simplify the analysis. We impose the following ansatz for circular strings in $\mathbb{R}\times S^5$ \cite{Frolov:2003qc}
\eq{
t=\kappa\tau\,,\quad\gamma=\gamma_0\,,\quad\theta=\frac{\pi}{4}\,,\quad\varphi_1=\omega\tau+m\sigma\,,\quad\varphi_2=\omega\tau-m\sigma\,,\quad\varphi_3=\nu\tau\,.
}
The equations of motion reduce to
\eq{
\omega^2=\nu^2+m^2\,.
}
The first Virasoro constraint reads
\eq{
\kappa^2=\omega^2-m^2\cos(2\gamma_0)\,,
}
while the second is automatically satisfied. The corresponding algebraic curve follows from
\eq{
j_S={\rm diag}\left\{\frac{i}{2}(2\omega+\nu),-\frac{i}{2}(2\omega-\nu),-\frac{i}{2}\nu,-\frac{i}{2}\nu\right\}
}
for the case of $\tau$-factorization. In the restriction $S^5\longrightarrow S^3$ we recover the curve in \eqref{s3circt'}. The spectral curve with imposed Virasoro constraints can be written in the following form
\begin{equation}
y^4+\frac{2\kappa^2z^2y^2}{(1-z^2)^2}+\frac{(\nu^2+2\omega^2)y^2}{2}+\frac{i\nu\omega^2(1+z^2)y}{1-z^2}+
\frac{4\kappa^4z^2(1+z^2)^2+\nu^2(\nu^2-4\omega^2)(1-z^4)^2}{16(1-z^2)^4}=0\,.
\end{equation}

\sect{Curves for rigid strings in ${\rm AdS}_5$}\label{sec6}

In this section we consider various algebraic curves corresponding to rigid spinning strings in AdS$_5$ \cite{Arutyunov:2003uj,Larsen:2003tb,Tirziu:2009ed}. The string shapes are either circular or folded, depending on the parameters of the solution. We will start with a straightforward generalization of the GKP string with ansatz
\eq{
t=\kappa\tau\,,\qquad\rho=\rho(\sigma)\,,\qquad\psi=\frac{\pi}{4}\,,\qquad\phi_1=w\tau\,,\qquad\phi_2=w\tau\,.
}
The equation of motion for $\rho$ is the same as the one in the GKP case, with a solution expressed in terms of the Jacobi amplitude function
\eq{
\rho(\sigma)={\bf am}\!\left(\sqrt{\kappa^2-w^2+c_1}\,(\sigma+c_2)\Bigg|\frac{\kappa^2-w^2}{\kappa^2-w^2+c_1}\right).
\label{ads5foldedsol}
}
We set $c_2=0$ by imposing that $\rho(0)=0$. The Virasoro constraints reduce to $c_1=w^2$. The components of the MC one-form can be easily computed, and clearly we have factorization only with respect to $\tau$. We obtain, using \eqref{eqcurrents}, that the constant current $j_S$ has the same form as the one for circular strings in AdS$_5$ \eqref{ads5circjs}. The resulting spectral curve is actually reducible. It decomposes to the curve
\eq{
y^2-iwy+\frac{\kappa^2}{4}+\frac{\kappa w}{2}\frac{1+z^2}{1-z^2}+\frac{(w^2-c_1)z^2}{(1-z^2)^2}=0
\label{ads5folded}
}
and a similar one, obtained from \eqref{ads5folded} by replacing $w$ with $-w$. Using the Virasoro constraints we obtain
\eq{
y^2-iwy+\frac{\kappa^2}{4}+\frac{\kappa w}{2}\frac{1+z^2}{1-z^2}=0\,.
\label{ads5foldedvir}
}
The reducibility of the curve can be traced to the fact that the rotation of the string in $S^3\subset{\rm AdS}_5$ can be considered as rotation along $S^1$, i.e. we have effective dynamics only in AdS$_3$.

Let us consider other non-trivial examples of rigid strings in AdS$_5$. Their algebraic curves are genus-3 in general. The first solution describes a circular string with equal spins $S_1=S_2=S$ in AdS$_5$. Starting with the following ansatz
\eq{
t=\kappa\tau\,,\qquad\rho=\rho(\sigma)\,,\qquad\psi=\psi(\sigma)\,,\qquad\phi_1=w\tau\,,\qquad\phi_2=w\tau\,,
\label{rigidsol}
}
we obtain from the equation of motion for $\psi$ and the first Virasoro constraint that
\eq{
\psi'=\frac{c}{\sinh^2\!\rho}\,,\qquad\rho'^2=\kappa^2\cosh^2\!\rho-\left(w^2+\frac{c^2}{\sinh^4\!\rho}\right)\sinh^2\!\rho\,.
}
The above expressions for the derivatives allow us to compute the corresponding spectral curve in the familiar fashion. The constant current is again as in \eqref{ads5circjs}, and the curve is
\eq{
y^4+\frac12(\kappa^2+2w^2)y^2+\frac{i\kappa w^2(1+z^2)}{1-z^2}y+\frac{\kappa^4}{16}-\frac{\kappa^2w^2(1+z^2)^2+4(\kappa^2-w^2)c^2z^2}{4(1-z^2)^2}=0\,.
\label{rigidcurve}
}
Note that when $c=0$ we recover \eqref{ads5foldedvir} as expected. Let us provide a number of limiting cases, which may be of interest to the reader. First, in the limit where the string shape approaches that of a round circle we have $c^2\rightarrow\frac{\kappa^4}{4(w^2-\kappa^2)}$, and
\eq{
y^4+\frac12(\kappa^2+2w^2)y^2+\frac{i\kappa w^2(1+z^2)}{1-z^2}y+\frac{\kappa^2(\kappa^2-4w^2)(1+z^2)^2}{16(1-z^2)^2}=0\,.
}
The large-spin limit $S/\sqrt{\lambda}\gg1$, which is satisfied by $\kappa=w\gg1$, of the string solution~\eqref{rigidsol} yields the algebraic curve
\eq{
y^4+\frac{3\kappa^2}{2}y^2+\frac{i\kappa^3(1+z^2)}{1-z^2}y-\frac{\kappa^4(3+z^2)(1+3z^2)}{16(1-z^2)^2}=0\,.
\label{largespin}
}
Another particularly interesting case is the circular winding string in AdS$_5$ with ansatz
\eq{
t=\kappa\tau\,,\qquad\rho=\rho_0\,,\qquad\psi=n\sigma\,,\qquad\phi_1=w\tau\,,\qquad\phi_2=w\tau\,,
}
where $n\in\mathbb{Z}$. The equation of motion for $\psi$ translates into $c=n\sinh^2\rho_0$, while the first Virasoro constraint gives $(w^2+n^2)\tanh^2\rho_0=\kappa^2$. All this leads to the following spectral curve
\eq{
y^4+\frac12(\kappa^2+2w^2)y^2+\frac{i\kappa w^2(1+z^2)}{1-z^2}y+\frac{\kappa^2(\kappa^2-4w^2)(1+z^2)^2}{16(1-z^2)^2}=0\,,
}
where we have also used that $w^2=\kappa^2+n^2$, which follows from the equation of motion for $\rho$. Similarly to the $S^3$ case the curve coincides with \eqref{ads5circ}, because the solutions are related to each other by a global SU(2) rotation of the $S^3$ part.

Finally, we will analyze the circular string with ansatz
\eq{
t=\kappa\tau\,,\qquad\rho=\rho(\sigma)\,,\qquad\psi=\psi(\sigma)\,,\qquad\phi_1=w\tau\,,\qquad\phi_2=\kappa\tau\,.
\label{rigidsol'}
}
This solution describes a string with two unequal (in general) spins $S_1\neq S_2$ in AdS$_5$. In order to obtain the algebraic curve, first we need to change the coordinates $(\rho,\psi)$ to $(\chi,\xi)$ defined accordingly
\eq{
\cosh\rho=\cosh\chi\cosh\xi\,,\qquad\sin\psi=\frac{\cosh\chi\sinh\xi}{\sqrt{\cosh^2\chi\sinh^2\xi+\sinh^2\chi}}\,.
}
Making use of the equation of motion for $\xi$ and the conformal constraints, we obtain
\eq{
\xi'=\frac{c}{\cosh^2\chi}\,,\qquad\chi'^2=\left(\kappa^2-\frac{c^2}{\cosh^4\chi}\right)\cosh^2\chi-w^2\sinh^2\chi\,.
}
After that we apply the procedure outlined in Section \ref{sec2}. The constant current is
\eq{
j_S={\rm diag}\left\{\frac{i}{2}(2\kappa+w),-\frac{iw}{2},-\frac{i}{2}(2\kappa-w),-\frac{iw}{2}\right\},
}
and the spectral curve is of the form
\eq{
y^4+\frac12(2\kappa^2+w^2)y^2+\frac{i\kappa^2w(1+z^2)}{1-z^2}y+\frac{w^4}{16}-\frac{\kappa^2w^2(1+z^2)^2+4(\kappa^2-w^2)c^2z^2}{4(1-z^2)^2}=0\,.
}
The large-spin limit, in which the spins become equal to each other ($S_1=S_2=S$) and $S/\sqrt{\lambda}\gg1$, is imposed through the condition $\kappa=w\gg1$ in \eqref{rigidsol'}. The algebraic curve that follows coincides with the one in \eqref{largespin}, because the two solutions are intimately related.

\sect{Conclusion}\label{sec7}

One of the main tools for comprehension of the inner mechanics of the AdS/CFT correspondence has been the classical integrability of the $\axs$ string. It allows the existence of a flat connection, on which we build an infinite number of conserved charges through its monodromies. The eigenvalues of the monodromy operator enter a particular Riemann surface known as spectral curve. The curve encodes all viable data about the string solution and paves the road for its semiclassical quantization. For these reasons and following the recent resurgence of interest in this field \cite{Janik:2012ws,Ryang:2012uf,Dekel:2013dy} we have constructed Lax operators and corresponding algebraic curves for an array of classical string solutions in $\axs$. In addition, we have studied various special cases. A possible continuation of this work that we intend to pursue is the extension of our results to other backgrounds such as AdS$_4\times CP^3$ and other classes of string solutions.

\section*{Acknowledgments}
The authors would like to thank H. Dimov for valuable discussions and careful reading of the paper. This work was supported in part by the Austrian Science Fund (FWF) project I 1030-N16.

\appendix

\sect{Reconstruction of string solutions from spectral curves in ${\rm AdS}_3$}

The author of \cite{Dekel:2013dy} provided a relatively simple procedure for the reconstruction of classical string solutions (up to conformal transformations and worldsheet translations) from their algebraic curves. Here we will rederive briefly the method, customizing it for our needs. The spectral curves in AdS$_3$ and $S^3$ have the universal form \eqref{curvepoly} with coefficients in the case of factorization with respect to $\tau$
\al{
c_0&=\tr[(j_\tau-j_S)^2]\,,\qquad c_1=2\tr[(j_\tau-j_S)j_\sigma]\,,\\
c_2&=\tr[2(j_\tau-j_S)j_S+j_\sigma^2]\,,\qquad c_3=2\tr(j_\sigma j_S)\,,\qquad c_4=\tr(j_S^2)\,.\nonumber
}
The starting point of the reconstruction is an ansatz for the worldsheet currents. Most of our curves have $c_4\neq0$, which leads to the following ansatz
\begin{equation}
j_\tau=S^{-1}\!\left(\begin{array}{cc}
\delta(\sigma) & \epsilon(\sigma)\\
\zeta(\sigma) & -\delta(\sigma)\end{array}\right)S\,,\quad
j_\sigma=S^{-1}\!\left(\begin{array}{cc}
\frac{c_3}{2\sqrt{2c_4}} & \beta(\sigma)\\
\gamma(\sigma) & -\frac{c_3}{2\sqrt{2c_4}}\end{array}\right)S\,,\quad
S=e^{\tau\sigma_3\sqrt{\frac{c_4}{2}}}.
\label{ads3ansatz}
\end{equation}
We can use the defining equations for $c_0,\,c_1,\,c_2$ to fix three of the functions in \eqref{ads3ansatz}, say $\delta$, $\epsilon$ and $\zeta$, in terms of the remaining two $\beta$ and $\gamma$
\al{
\delta&=\frac{4c_4(c_2+2c_4)-c_3^2-8c_4\beta\gamma}{8\sqrt{2}\,c_4^{3/2}}\,,\qquad\epsilon=\frac{c_0-c_4+2\sqrt{2}\,\sqrt{c_4}\delta-2\delta^2}{2\zeta}\,,\\
\zeta&=\frac{8c_1c_4^2+c_3^3-4c_3c_4(c_2-2\beta\gamma)}{32c_4^2\beta}\pm\frac{1}{32c_4^2\beta}\{8c_4\beta\gamma[-64c_0c_4^3-16c_2c_3^2c_4+3c_3^4\nonumber\\
&+16c_4^2(c_1c_3+c_2^2)+8c_4\beta\gamma(-8c_2c_4+3c_3^2+8c_4\beta\gamma)]+(8c_1c_4^2-4c_2c_3c_4+c_3^3)^2\}^{1/2}.
}
Inserting these functions in the equations of motion and the MC equations leaves us with two first-order coupled nonlinear differential equations for $\beta$ and $\gamma$
\al{
\beta'&=-\frac{8c_1c_4^2+c_3^3-4c_3c_4(c_2-2\beta\gamma)}{16\sqrt{2}c_4^{3/2}\gamma}+\frac{1}{16\sqrt{2}c_4^{3/2}\gamma}\{8c_4\beta\gamma[-64c_0c_4^3-16c_2c_3^2c_4+3c_3^4\nonumber\\
&+16c_4^2(c_1c_3+c_2^2)+8c_4\beta\gamma(-8c_2c_4+3c_3^2+8c_4\beta\gamma)]+(8c_1c_4^2-4c_2c_3c_4+c_3^3)^2\}^{1/2},
}
\al{
\gamma'&=\frac{8c_1c_4^2+c_3^3-4c_3c_4(c_2-2\beta\gamma)}{16\sqrt{2}c_4^{3/2}\beta}+\frac{1}{16\sqrt{2}c_4^{3/2}\beta}\{8c_4\beta\gamma[-64c_0c_4^3-16c_2c_3^2c_4+3c_3^4\nonumber\\
&+16c_4^2(c_1c_3+c_2^2)+8c_4\beta\gamma(-8c_2c_4+3c_3^2+8c_4\beta\gamma)]+(8c_1c_4^2-4c_2c_3c_4+c_3^3)^2\}^{1/2}.
}
In general the equations can be solved in terms of elliptic functions, but the expressions are rather complicated, and so we will analyze only particular simple cases. For example, if $c_1=c_3=0$, we obtain that $\gamma(\sigma)=\gamma_1\beta(\sigma)$ with $\gamma_1={\rm const}$.

After we have obtained the currents $j_\tau$ and $j_\sigma$ we set the following ansatz for the group element\footnote{We have assumed that $c_0\neq0$, which conforms to the curves obtained in the present paper.}
\eq{
g=e^{\tau\sigma_3\sqrt{\frac{c_0}{2}}}\left(\begin{array}{cc}
a(\sigma) & b(\sigma)\\
c(\sigma) & d(\sigma)\end{array}\right)S\,.
}
There are actually only three independent functions to be found because of the condition $\det g=1$. The equation $j_\tau\equiv g^{-1}\partial_\tau g$ gives a set of algebraic equations, from which we can express two of the functions in terms of the third. Then the relation $j_\sigma\equiv g^{-1}\partial_\sigma g$ provides a first-order differential equation for the last remaining function. Thus one can reconstruct the initial string solution. Note, however, that in general there are severe computational difficulties in carrying out the above procedure. For this reason we will only provide as examples the expressions for $\beta$ and $\gamma$. First, in the case of circular strings with solution~\eqref{circsola} the corresponding curve is given by \eqref{curvecirctau'}. Using the machinery described above, we find that
\al{
\beta&=\frac{\gamma_1e^{-ik(\sigma+\gamma_2)}}{\eta}\,,\qquad\eta=\cosh\rho_0-i\tan[k(\sigma+\gamma_2)\cosh\rho_0]\,,\\ \gamma&=\gamma_1^{-1}k^2\cosh^2\rho_0e^{ik(\sigma+\gamma_2)}|\eta|^2\eta\,.
}
Roughly speaking, the dependence on $\sigma$ is exponential, which is expected of circular strings.

The second example is the folded spinning string satisfying the Virasoro constraints, with curve given in eq. (6.31) of \cite{Dekel:2013dy}. The expressions for $\beta$ and $\gamma$ are
\eq{
\gamma(\sigma)=\gamma_1\beta(\sigma)=\frac{\sqrt{\gamma_1}\!\left(\kappa^2-\omega^2+\omega^2\,{\bf sn}\!\left(\sqrt{\kappa^2-\omega^2}\,(\sigma+\gamma_2),-\frac{\omega^2}{\kappa^2-\omega^2}\right)^2\right)^{\!1/2}}
{{\bf sn}\!\left(\sqrt{\kappa^2-\omega^2}\,(\sigma+\gamma_2),-\frac{\omega^2}{\kappa^2-\omega^2}\right)}\,.
}
They are given in terms of Jacobi elliptic functions as expected.

\sect{Reconstruction of string solutions from spectral\\ curves in $S^3$}

The analysis of the $S^3$ case is completely analogous to the previous one. For $c_4\neq0$ we may start with the following ansatz
\begin{equation}
j_\tau=S^{-1}\!\left(\!\!\begin{array}{cc}
i\delta(\sigma) &\!\! \xi(\sigma)+i\eta(\sigma)\\
-\xi(\sigma)+i\eta(\sigma) &\!\! -i\delta(\sigma)\end{array}\!\!\right)S\,,\
j_\sigma=S^{-1}\!\left(\!\!\begin{array}{cc}
\frac{c_3}{2\sqrt{2c_4}} &\!\! \beta(\sigma)+i\gamma(\sigma)\\
-\beta(\sigma)+i\gamma(\sigma) &\!\! -\frac{c_3}{2\sqrt{2c_4}}\end{array}\!\!\right)S\,,
\end{equation}
where $S=\exp\!\big(\tau\sigma_3\sqrt{c_4/2}\big)$. Again we can express three of the functions, say $\delta$, $\xi$ and $\eta$, in terms of $\beta$ and $\gamma$. With the help of the equations of motion and the MC equations we obtain two first-order nonlinear coupled differential equations for $\beta$ and $\gamma$
\al{
\beta'&=-\frac{i\gamma[8c_1c_4^2+c_3^3-4c_3c_4(c_2+2\alpha^2)]}{16\sqrt{2}c_4^{3/2}\alpha^2}+
\frac{i\beta{\rm sign}(\gamma)}{16\sqrt{2}c_4^{3/2}\alpha^2}\{8c_4\alpha^2[-64c_0c_4^3-16c_2c_3^2c_4+3c_3^4\nonumber\\
&+16c_4^2(c_1c_3+c_2^2)+ 8c_4\alpha^2(8c_2c_4-3c_3^2+8c_4\alpha^2)]-(8c_1c_4^2-4c_2c_3c_4+c_3^3)^2\}^{1/2},
}
\al{
\gamma'&=\frac{i\beta[8c_1c_4^2+c_3^3-4c_3c_4(c_2+2\alpha^2)]}{16\sqrt{2}c_4^{3/2}\alpha^2}+
\frac{i\gamma{\rm sign}(\gamma)}{16\sqrt{2}c_4^{3/2}\alpha^2}\{8c_4\alpha^2[-64c_0c_4^3-16c_2c_3^2c_4+3c_3^4\nonumber\\
&+16c_4^2(c_1c_3+c_2^2)+ 8c_4\alpha^2(8c_2c_4-3c_3^2+8c_4\alpha^2)]-(8c_1c_4^2-4c_2c_3c_4+c_3^3)^2\}^{1/2},
}
where $\alpha^2=\beta^2+\gamma^2$. Again in the special case of $c_1=c_3=0$ we find the simple relation $\gamma(\sigma)=\gamma_1\beta(\sigma)$.

After obtaining the currents $j_\tau$ and $j_\sigma$ we utilize the following ansatz for the group element
\eq{
g=e^{\tau\sigma_3\sqrt{\frac{c_0}{2}}}\left(\begin{array}{cc}
a(\sigma)+ib(\sigma) & c(\sigma)+id(\sigma)\\
-c(\sigma)+id(\sigma) & a(\sigma)-ib(\sigma)\end{array}\right)S\,.
}
Again because of the $\det g=1$ condition there are only three real independent functions to be found. The $j_\tau\equiv g^{-1}\partial_\tau g$ equation provides a set of algebraic equations, from which we can express two of the functions in terms of the third. Finally, the relation $j_\sigma\equiv g^{-1}\partial_\sigma g$ gives a first-order differential equation for the last remaining function. The functions $\beta$ and $\gamma$ for the example case of folded spinning string in $S^3$ with spectral curve \eqref{foldeds3} can be obtained via the above procedure as
\eq{
\gamma(\sigma)=-\gamma_1\beta(\sigma)=\gamma_1\sqrt{\frac{\kappa^2-\omega_1^2}{1+\gamma_1^2}}\,
{\bf sn}\!\left(\sqrt{\omega_2^2-\kappa^2}\,(\sigma+\gamma_2),\frac{\kappa^2-\omega_1^2}{\kappa^2-\omega_2^2}\right).
}


\begin{thebibliography}{99}

\bibitem{Maldacena:1997re}
  J.~M.~Maldacena,
  ``The large $N$ limit of superconformal field theories and supergravity,''
  Adv. Theor. Math. Phys. {\bf 2}, 231 (1998)
  [hep-th/9711200].

\bibitem{Gubser:1998bc}
  S.~S.~Gubser, I.~R.~Klebanov and A.~M.~Polyakov,
  ``Gauge theory correlators from non-critical string theory,''
  Phys. Lett. B {\bf 428}, 105 (1998)
  [hep-th/9802109].

\bibitem{Witten:1998qj}
  E.~Witten,
  ``Anti-de Sitter space and holography,''
  Adv. Theor. Math. Phys. {\bf 2}, 253 (1998)
  [hep-th/9802150].

\bibitem{Minahan:2002ve}
  J.~A.~Minahan and K.~Zarembo,
  ``The Bethe-ansatz for ${\cal N}=4$ super Yang-Mills,''
  JHEP {\bf 0303}, 013 (2003)
  [hep-th/0212208].

\bibitem{Bena:2003wd}
  I.~Bena, J.~Polchinski and R.~Roiban,
  ``Hidden symmetries of the $\axs$ superstring,''
  Phys. Rev. D {\bf 69}, 046002 (2004)
  [hep-th/0305116].

\bibitem{Frolov:2003qc}
  S.~Frolov and A.~A.~Tseytlin,
  ``Multispin string solutions in $\axs$,''
  Nucl. Phys. B {\bf 668}, 77 (2003)
  [hep-th/0304255].

\bibitem{Kazakov:2004qf}
  V.~Kazakov, A.~Marshakov, J.~Minahan and K.~Zarembo,
  ``Classical/quantum integrability in AdS/CFT,''
  JHEP {\bf 0405}, 024 (2004)
  [hep-th/0402207].

\bibitem{Kazakov:2004nh}
  V.~Kazakov and K.~Zarembo,
  ``Classical/quantum integrability in non-compact sector of AdS/CFT,''
  JHEP {\bf 0410}, 060 (2004)
  [hep-th/0410105].

\bibitem{Tseytlin:2010jv}
  A.~A.~Tseytlin,
  ``Review of AdS/CFT Integrability, Chapter II.1: Classical $\axs$ string solutions,''
  Lett. Math. Phys. {\bf 99}, 103 (2012)
  [arXiv:1012.3986].

\bibitem{Babelon:2003}
  O.~Babelon, D.~Bernard and M.~Talon,
  {\it Introduction to Classical Integrable Systems},
  Cambridge University Press, 2003.

\bibitem{Arutyunov:2009ga}
  G.~Arutyunov and S.~Frolov,
  ``Foundations of the $\axs$ Superstring. Part I,''
  J. Phys. A {\bf 42}, 254003 (2009)
  [arXiv:0901.4937].

\bibitem{Beisert:2010jr}
  N.~Beisert, C.~Ahn, L.~F.~Alday, Z.~Bajnok, J.~M.~Drummond, L.~Freyhult, N.~Gromov and R.~A.~Janik {\it et al.},
  ``Review of AdS/CFT Integrability: An Overview,''
  Lett. Math. Phys. {\bf 99}, 3 (2012)
  [arXiv:1012.3982].


\bibitem{Beisert:2004ag}
  N.~Beisert, V.~Kazakov, and K.~Sakai,
  ``Algebraic curve for the SO(6) sector of AdS/CFT,''
  Commun. Math. Phys. {\bf 263}, 611 (2006)
  [hep-th/0410253].

\bibitem{Beisert:2005bm}
  N.~Beisert, V.~A.~Kazakov, K.~Sakai and K.~Zarembo,
  ``The Algebraic curve of classical superstrings on $\axs$,''
  Commun. Math. Phys. {\bf 263}, 659 (2006)
  [hep-th/0502226].

\bibitem{Dorey:2006zj}
  N.~Dorey and B.~Vicedo,
  ``On the dynamics of finite-gap solutions in classical string theory,''
  JHEP {\bf 0607}, 014 (2006)
  [hep-th/0601194].

\bibitem{Dorey:2006mx}
  N.~Dorey and B.~Vicedo,
  ``A Symplectic Structure for String Theory on Integrable Backgrounds,''
  JHEP {\bf 0703}, 045 (2007)
  [hep-th/0606287].

\bibitem{SchaferNameki:2010jy}
  S.~Schafer-Nameki,
  ``Review of AdS/CFT integrability, Chapter II.4: The spectral curve,''
  Lett. Math. Phys. {\bf 99}, 169 (2012)
  [arXiv:1012.3989].

\bibitem{Sakai:2006bp}
  K.~Sakai and Y.~Satoh,
  ``A large spin limit of strings on $\axs$ in a non-compact sector,''
  Phys. Lett. B {\bf 645}, 293 (2007)
  [hep-th/0607190].

\bibitem{Zarembo:2010yz}
  K.~Zarembo,
  ``Algebraic Curves for Integrable String Backgrounds''
  [arXiv:1005.1342].

\bibitem{Alday:2005gi}
  L.~F.~Alday, G.~Arutyunov and A.~A.~Tseytlin,
  ``On integrability of classical superstrings in $\axs$,''
  JHEP {\bf 0507}, 002 (2005)
  [hep-th/0502240].

\bibitem{Gromov:2007aq}
  N.~Gromov and P.~Vieira,
  ``The $\axs$ superstring quantum spectrum from the algebraic curve,''
  Nucl. Phys. B {\bf 789}, 175 (2008)
  [hep-th/0703191].

\bibitem{Gromov:2011de}
  N.~Gromov, D.~Serban, I.~Shenderovich and D.~Volin,
  ``Quantum folded string and integrability: From finite size effects to Konishi dimension,''
  JHEP {\bf 1108}, 046 (2011)
  [arXiv:1102.1040].

\bibitem{Gromov:2013pga}
  N.~Gromov, V.~Kazakov, S.~Leurent and D.~Volin,
  ``Quantum Spectral Curve for Planar ${\cal N}=4$ Super-Yang-Mills Theory,''
  Phys. Rev. Lett. {\bf 112}, 011602 (2014)
  [arXiv:1305.1939].

\bibitem{Ishizeki:2011bf}
  R.~Ishizeki, M.~Kruczenski and S.~Ziama,
  ``Notes on Euclidean Wilson loops and Riemann Theta functions,''
  Phys. Rev. D {\bf 85}, 106004 (2012)
  [arXiv:1104.3567].

\bibitem{Janik:2012ws}
  R.~A.~Janik and P.~Laskos-Grabowski,
  ``Surprises in the AdS algebraic curve constructions -- Wilson loops and correlation functions,''
  Nucl. Phys. B {\bf 861}, 361 (2012)
  [arXiv:1203.4246].

\bibitem{Ryang:2012uf}
  S.~Ryang,
  ``Algebraic Curves for Long Folded and Circular Winding Strings in $\axs$,''
  JHEP {\bf 1302}, 107 (2013)
  [arXiv:1212.6109].

\bibitem{Dekel:2013dy}
  A.~Dekel,
  ``Algebraic Curves for Factorized String Solutions,''
  JHEP {\bf 1304}, 119 (2013)
  [arXiv:1302.0555].


\bibitem{Arutyunov:2003uj}
  G.~Arutyunov, S.~Frolov, J.~Russo and A.~A.~Tseytlin,
  ``Spinning strings in $\axs$ and integrable systems,''
  Nucl. Phys. B {\bf 671}, 3 (2003)
  [hep-th/0307191].

\bibitem{Arutyunov:2003rg}
  G.~Arutyunov and M.~Staudacher,
  ``Matching higher conserved charges for strings and spins,''
  JHEP {\bf 0403}, 004 (2004)
  [hep-th/0310182].

\bibitem{Arutyunov:2003za}
  G.~Arutyunov, J.~Russo and A.~A.~Tseytlin,
  ``Spinning strings in $\axs$: New integrable system relations,''
  Phys. Rev. D {\bf 69}, 086009 (2004)
  [hep-th/0311004].

\bibitem{Frolov:2003xy}
  S.~Frolov and A.~A.~Tseytlin,
  ``Rotating string solutions: AdS/CFT duality in non-supersymmetric sectors,''
  Phys. Lett. B {\bf 570}, 96 (2003)
  [hep-th/0306143].

\bibitem{Alday:2005ww}
  L.~F.~Alday, G.~Arutyunov and S.~Frolov,
  ``Green-Schwarz Strings in TsT-transformed backgrounds,''
  JHEP {\bf 0606}, 018 (2006)
  [hep-th/0512253].

\bibitem{Larsen:2003tb}
  A.~L.~Larsen and A.~Khan,
  ``Novel explicit multi spin string solitons in AdS$_5$,''
  Nucl. Phys. B {\bf 686}, 75 (2004)
  [hep-th/0312184].

\bibitem{Tirziu:2009ed}
  A.~Tirziu and A.~A.~Tseytlin,
  ``Semiclassical rigid strings with two spins in AdS$_5$,''
  Phys. Rev. D {\bf 81}, 026006 (2010)
  [arXiv:0911.2417].

\end{thebibliography}
\end{document}